\documentclass[letterpaper,journal]{IEEEtran}
\usepackage{amsmath,amsfonts}
\usepackage{algorithmic}
\usepackage{algorithm}
\usepackage{array}
\usepackage{textcomp}
\usepackage{stfloats}
\usepackage{url}
\usepackage{verbatim}
\usepackage{graphicx}
\usepackage{subcaption}
\usepackage{cite}
\usepackage{bm}
\usepackage{amsthm,amssymb}
\usepackage{mathrsfs}
\usepackage{soul,color,xcolor}
\usepackage{epstopdf}

\allowdisplaybreaks[4]
\def\BibTeX{{\rm B\kern-.05em{\sc i\kern-.025em b}\kern-.08em
		T\kern-.1667em\lower.7ex\hbox{E}\kern-.125emX}}
\usepackage{balance}
\begin{document}
%	\title{Rotatable Antenna-Enabled Dynamic Air-Ground Integrated Communications: TDMA vs NOMA}
	\title{Energy Efficiency Optimization for Rotatable Antenna-Enabled Uplink NOMA Systems}
	\author{Yixuan Li, Jun Wang, Hongbo Xu, and Ji Wang, \IEEEmembership{Senior Member,~IEEE}
		\thanks{\textit{(Corresponding author: Jun Wang.)}}
		\thanks{Yixuan Li, Jun Wang, Hongbo Xu, and Ji Wang are with the Department of Electronics and Information Engineering, Central China Normal University, Wuhan 430079, China (e-mail: yixuanli@mails.ccnu.edu.cn; junwang@ccnu.edu.cn; xuhb@ccnu.edu.cn; jiwang@ccnu.edu.cn).		
	}}
	
	\maketitle
	
	\begin{abstract}
		This paper investigates a rotatable antenna (RA)-enabled uplink non-orthogonal multiple access (NOMA) system, where a base station equipped with multiple independently RAs serves both ground and aerial users. Specifically, we formulate an energy efficiency (EE) maximization problem by jointly optimizing receive beamforming, user power allocation, and RA rotation. To make the problem tractable, a new block coordinate descent-based algorithm is developed, in which the receive beamforming is updated via the minimum mean square error criterion, while the power allocation and RA rotation are handled by fractional programming and successive convex approximation. Numerical results demonstrate the EE superiority of the proposed RA-NOMA scheme over several benchmarks.
	\end{abstract}
	

	\section{Introduction}
	Non-orthogonal multiple access (NOMA) has attracted considerable attention as an effective multiple access paradigm capable of enhancing spectral efficiency and accommodating massive connections in future wireless networks \cite{7676258}. Considerable research efforts have therefore been devoted to integrating NOMA with channel-reconfiguration technologies to improve channel conditions and alleviate severe path loss. For instance, reconfigurable intelligent surfaces (RISs) \cite{9133094} assist NOMA by establishing additional reflected links to enhance signal propagation, while movable antennas (MAs) \cite{10535440} and fluid antennas \cite{10855346} exploit position reconfiguration to reshape channels. More recently, pinching antennas (PAs) \cite{11131179} have been investigated to shorten the propagation distance between antennas and users by creating strong line-of-sight (LoS) links, whereas six-dimensional (6D)-MA in \cite{11142311} further extends the channel reconfiguration capability by jointly exploiting antenna displacement and array-plane rotation.
	
	In this context, an rotatable antenna (RA) technology has attracted growing research interest due to its ability to improve wireless transmission performance through flexible adjustment of antenna orientations \cite{zheng2026rotatable,11222668,11427014,11098736,11347570,11039664}. In contrast to conventional fixed-antenna (FA) architectures, each RA can adaptively rotate its boresight toward a desired spatial direction, thereby combining the high directional gain of directional antennas with additional spatial degrees of freedom (DoFs) enabled by antenna rotation \cite{zheng2026rotatable}. Such a capability allows the received signal power to be effectively strengthened while undesired radiation directions can be better suppressed. Compared with existing channel-reconfiguration technologies \cite{9133094,10535440,10855346,11131179,11142311}, RA can offer a relatively compact and implementation-friendly solution. It does not require additional surface deployment as in RIS-assisted systems, nor does it rely on the relatively sophisticated structural designs needed by MAs, fluid antennas, and 6D-MAs. Moreover, unlike PAs, RA avoids large-scale waveguide deployment and the associated waveguide propagation loss. Instead, RA can be practically realized through simple rotation mechanisms, such as compact servo motors, making it a lightweight yet effective approach for antenna-side channel reconfiguration \cite{zheng2026rotatable,11222668,11427014}. Recent works have verified the effectiveness of RA technology in a wide range of wireless applications, including physical-layer security \cite{11098736}, cognitive radio \cite{11347570}, and integrated sensing and communications \cite{11039664}. Despite these advances, the integration of RA technology with NOMA remains largely unexplored. To the best of our knowledge, this is the first work to investigate an RA-enabled NOMA system, where RA-induced spatial DoFs are exploited to enhance signal reception and mitigate multiuser interference for successive interference cancellation (SIC) decoding.
	
	In this paper, we investigate a RA-assisted uplink NOMA system, where a BS equipped with multiple independently rotatable antennas simultaneously receives information from multiple ground and aerial users. To balance the achievable rate and power consumption, we formulate an energy efficiency (EE) maximization problem by jointly optimizing the BS receive beamforming, user transmit power, and RA rotation angles. The formulated problem is highly non-convex due to the coupled optimization variables and the nonlinear RA directional gain. To address this challenge, we develop a new block coordinate descent (BCD)-based algorithm. Specifically, the receive beamforming vectors are obtained based on the minimum mean square error (MMSE) criterion, an fractional programming (FP)-successive convex approximation (SCA) algorithm is developed for user power allocation, and the RA rotation angles are optimized through SCA and convex approximation. Simulation results demonstrate that the proposed RA-NOMA scheme significantly outperforms the RA-space division multiple access (SDMA), RA-time division multiple access (TDMA), isotropic-antenna (IA)-NOMA, and fixed-antenna (FA)-NOMA schemes.

	\section{System Model and Problem Formulation}
	\begin{figure}[!t]
		\centering
		\includegraphics[width=3.2in]{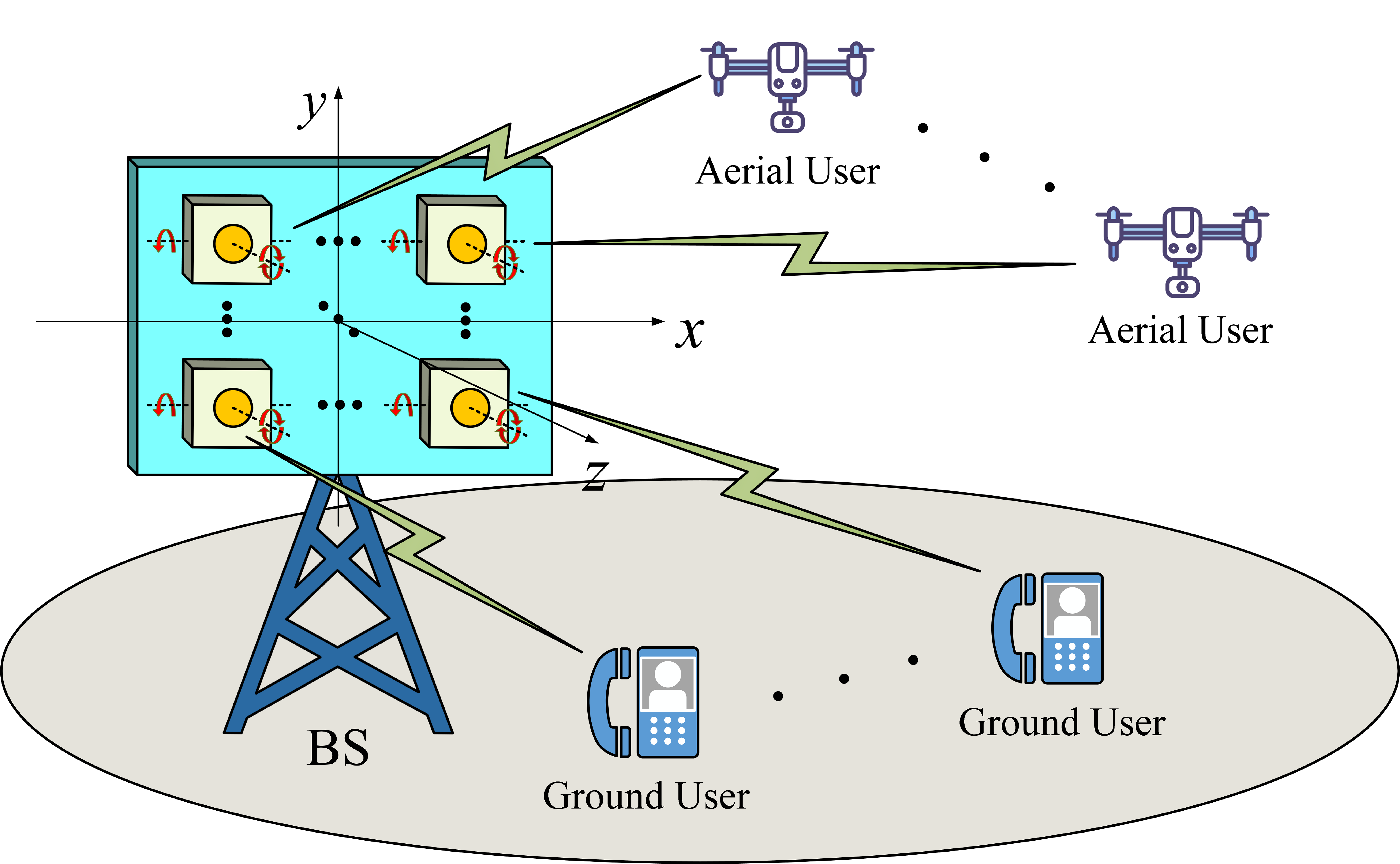}
		\caption{RA enabled uplink NOMA system.}
		\label{system_model}
	\end{figure}
	In Fig. \ref{system_model}, an uplink air-ground integrated communication system is illustrated, where a BS equipped with $N = N_x \times N_y$ RAs simultaneously receives information signals from $K = K_G+K_A$ single-antenna mobile users, including $K_G$ ground users and $K_A$ aerial users. Here, $N_x$ and $N_y$ denote the numbers of rows and columns of the RA array, respectively. To facilitate the subsequent mathematical formulation, we use ${\cal N} \buildrel \Delta \over = \left\{1,...,N\right\}$ and ${\cal K} \buildrel \Delta \over = \left\{1,...,K\right\}$ to represent the index sets of the RAs and users, respectively. We assume that the considered system is described in a three-dimensional (3D) Cartesian coordinate system with the center of the RA array as the origin, where the RA array is placed on the $x$-$y$ plane, and the outward normal of its front face is aligned with the positive $z$-axis. A minimum spacing $\Delta$ is imposed between any two RAs to reduce mutual coupling. Therefore, the coordinate of the $n$-th RA is denoted by ${\bf z}_n = {\bf z}_{n_x,n_y} = \left[n_x \Delta,n_y \Delta,0\right]^T$. Specifically, when $N_x$ and $N_y$ are odd, we set ${n_x} \in \left\{ {0, \pm 1,..., \pm \frac{{{N_x} - 1}}{2}} \right\}$ and ${n_y} \in \left\{ {0, \pm 1,..., \pm \frac{{{N_y} - 1}}{2}} \right\}$; when $N_x$ and $N_y$ are even, we set ${n_x} \in \left\{ {\pm 0.5,..., \pm \frac{{{N_x} - 1}}{2}} \right\}$ and ${n_y} \in \left\{ {\pm 0.5,..., \pm \frac{{{N_y} - 1}}{2}} \right\}$. Furthermore, the coordinate of the $k$-th user is denoted by ${\bf u}_k = \left[x_{{\rm U},k},y_{{\rm U},k},z_{{\rm U},k}\right]^T$. When $y_{{\rm U},k} = -H_{\rm BS}$, the $k$-th user is identified as a ground user, where $H_{\rm BS}$ denotes the height of the RA array center. 
	
	We assume that each RA can rotate independently in the considered system. Accordingly, the pointing vector of the $n$-th RA is denoted by ${\vec{\bf{f}}_n} = {\left[ {{f_{x,n}},{f_{y,n}},{f_{z,n}}} \right]^T} \in {{\mathbb R}^{3 \times 1}}$, where ${f_{x,n}} = \sin {\theta _{z,n}}\cos {\varphi _{a,n}}$, ${f_{y,n}} = \sin {\theta _{z,n}}\sin {\varphi _{a,n}}$, and ${f_{z,n}} = \cos {\theta _{z,n}}$ denote the $x$-, $y$-, and $z$-axis components of $\vec{\bf f}_n$, respectively. $\theta _{z,n} \in \left[0,\theta_{\max}\right]$ and $\varphi _{a,n} \in \left[0,2 \pi\right]$ denote the zenith and azimuth angles of the $n$-th RA, respectively, where $\theta _{\max} \in \left[0,\frac{\pi}{2}\right]$ represents the maximum zenith angle supported by each RA. Thus, the pointing vector ${\vec{\bf f}_n}$ should be satisfied
	\begin{flalign}
		\big\| {{\vec{\bf{f}}_n}} \big\| = 1,~0 \le \arccos \big( {\vec{\bf{f}}_n^T{\bf{e}}} \big) \le {\theta _{\max }},\forall n, \label{c_f}
	\end{flalign}
	where ${\bf e} = \left[0,0,1\right]^T$. Since the channel gain between the BS and each user depends on the incident angle at the RAs, a unified directional gain model is adopted to characterize the angle-dependent channel variation. Thus, the directional gain from the $k$-th user to the $n$-th RA is given by ${G_{n,k}} = {G_0}{\left( {\vec{\bf{f}}_n^T{\vec{\bf{q}}_{n,k}}} \right)^{2p}}$, where ${\vec{\bf{q}}_{n,k}} = \frac{{{{\bf{u}}_k} - {{\bf{z}}_n}}}{{\left\| {{{\bf{u}}_k} - {{\bf{z}}_n}} \right\|}}$. Here, $p$ and $G_0 = 2\left(2p+1\right)$ denote the directivity factor and the maximum directional gain of the RA, respectively. According to \cite{11427014,11098736,11347570}, the channel coefficient from the $n$-th RA to the $k$-th user is expressed as{\footnote{For a high-precision performance comparison, we consider line-of-sight (LoS)-dominated channels between the BS and all users, so as to isolate the impact of the multiple access strategy from the random fluctuations caused by non-line-of-sight (NLoS) propagation. Nevertheless, the proposed algorithms are generally applicable and can also be employed in system models with NLoS components.}}
	\begin{flalign}
		{h_{n,k}} = \sqrt {\frac{{A{G_{n,k}}}}{{4\pi d_{n,k}^2}}} {e^{ - j\frac{{2\pi }}{\lambda }{d_{n,k}}}} = {\bar g _{n,k}}{\left( {\vec {\bf{f}} _n^T{{\vec {\bf{q}} }_{n,k}}} \right)^p},
	\end{flalign}
	where ${\bar g _{n,k}} = \sqrt {\frac{{A{G_0}}}{{4\pi d_{n,k}^2}}} {e^{ - j\frac{{2\pi }}{\lambda }{d_{n,k}}}}$. $d_{n,k} = \left\|{\bf z}_n - {\bf u}_k\right\|$ denotes the distance between the $n$-th RA and the $k$-th user, while $\lambda$ and $A$ represent the carrier wavelength and the physical aperture size of the antenna, respectively. Accordingly, the channel from the BS to the $k$-th user is ${\bf h}_k = \left[h_{1,k},...,h_{N,k}\right]^T$.
	
	To exploit the spatial flexibility of RAs while enabling efficient simultaneous uplink transmission from multiple users, a NOMA protocol is adopted in the considered system. Therefore, the achievable spectral efficiency of the $k$-th user is
	\begin{flalign}
		R_k =& {\log _2}\left( {1 + \frac{{{P_k}{{\left| {{\bf{w}}_k^H{{\bf{h}}_k}} \right|}^2}}}{{\sum\nolimits_{i = k+1}^K {{P_i}{{\left| {{\bf{w}}_k^H{{\bf{h}}_i}} \right|}^2}}  + {\sigma ^2}}}} \right),
	\end{flalign}
	where ${\bf w}_k \in {\mathbb C}^{N \times 1}$ denotes the BS receive beamforming vectors. Here, $P_k$ denotes the transmit power of user $k$, while $\sigma^2$ is the noise power. In uplink NOMA transmission, SIC is employed at the BS for signal detection, where users are successively decoded according to the descending order of their effective channel gains.
	
	In this work, the system EE is maximized while guaranteeing the minimum QoS requirement $R_{\min}$ of each user. Accordingly, the corresponding problem is formulated as
	\begin{subequations}\label{p1}
		\begin{align}
			&{\rm{  }}\mathop {{\rm{max}}}\limits_{\left\{ {{{\bf{w}}_k}} \right\},\left\{P_k\right\},\left\{ {{\vec{\bf{f}}_n}} \right\}} \frac{\sum\nolimits_{k = 1}^K {R_k} }{\sum\nolimits_{k = 1}^K {P_k + P_{c}}} \\
			&{\rm{s}}{\rm{.t}}{\rm{.}}~R_k \ge {R_{\min }},~0 \le {P_k} \le P_{\max},\forall k,~\left( {\rm \ref{c_f}} \right),
		\end{align}
	\end{subequations}
	where $P_{c}$ represents the total constant circuit power.
	
	\section{The Proposed Algorithm}
	The non-concavity of the objective function, together with the non-convex feasible region, renders problem~(\ref{p1}) difficult to solve directly. To circumvent this difficulty, we decompose the original problem into several tractable subproblems.
	
	\subsection{Beamforming Optimization}
	Given $\left\{P_k\right\}$ and $\big\{\vec{\bf f}_n\big\}$, the receive beamforming vector for each user can be obtained by the MMSE criterion. Specifically, after the signals of users $1,...,k-1$ have been successfully decoded and removed by SIC, the remaining interference for decoding user $k$ comes from users $k+1,...,K$. Accordingly, the MMSE receive beamforming vector for the $k$-th user is given by
	\begin{flalign}
		{\bf{w}}_k^{\star} = \frac{{{{\left( {\sum\nolimits_{i = k + 1}^K {{P_i}{{\bf{h}}_i}{\bf{h}}_i^H}  + {\sigma ^2}{\bf{I}}} \right)}^{ - 1}}{{\bf{h}}_k}}}{{\left\| {{{\left( {\sum\nolimits_{i = k + 1}^K {{P_i}{{\bf{h}}_i}{\bf{h}}_i^H}  + {\sigma ^2}{\bf{I}}} \right)}^{ - 1}}{{\bf{h}}_k}} \right\|}}. \label{solvew}
	\end{flalign}
	
	\subsection{Power Allocation}
	Given $\left\{{\bf w}_k\right\}$ and $\big\{\vec{\bf f}_n\big\}$, problem (\ref{p1}) can be rewritten as
	\begin{subequations}\label{p2}
		\begin{align}
			&{\rm{  }}\mathop {{\rm{max}}}\limits_{\left\{P_k\right\}} \frac{\sum\nolimits_{k = 1}^K {R_k} }{\sum\nolimits_{k = 1}^K {P_k + P_{c}}} \\
			&{\rm{s}}{\rm{.t}}{\rm{.}}~R_k \ge {R_{\min }},~0 \le {P_k} \le P_{\max},\forall k.
		\end{align}
	\end{subequations}
	To handle the non-convex fractional problem (\ref{p2}), we apply the FP method and reformulate it as
	\begin{subequations}\label{p3}
		\begin{align}
			{\rm{  }}\mathop {{\rm{max}}}\limits_{\left\{P_k\right\}} 
			&\sum\limits_{k = 1}^K {{{\log }_2}\left( {1 + \frac{{{P_k}{{\left| {{\bf{w}}_k^H{{\bf{h}}_k}} \right|}^2}}}{{\sum\nolimits_{i = k + 1}^K {{P_i}{{\left| {{\bf{w}}_k^H{{\bf{h}}_i}} \right|}^2}}  + {\sigma ^2}}}} \right)}  \notag\\
			&- \eta  {\left(\sum\nolimits_{k = 1}^K {{P_k}}  + {P_{c}} \right) }\\
			{\rm{s}}{\rm{.t}}{\rm{.}}~&{P_k}{\left| {{\bf{w}}_k^H{{\bf{h}}_k}} \right|^2} \ge \left( {{2^{{R_{\min }}}} - 1} \right) \left( {\sum\limits_{i = k + 1}^K {{P_i}\left| {{\bf{w}}_k^H{{\bf{h}}_i}} \right|^2}  + {\sigma ^2}} \right),\notag\\
			&0 \le {P_k} \le P_{\max},\forall k, \label{p3-b}
		\end{align}
	\end{subequations}
	where $\eta$ denotes the Dinkelbach parameter, which is iteratively updated as
	\begin{flalign}
		{\eta ^{\left( l \right)}} = \frac{{\sum\limits_{k = 1}^K {{{\log }_2}\left( {1 + \frac{{P_k^{\left( l \right)}{{\left| {{\bf{w}}_k^H{{\bf{h}}_k}} \right|}^2}}}{{\sum\nolimits_{i = k + 1}^K {P_i^{\left( l \right)}{{\left| {{\bf{w}}_k^H{{\bf{h}}_i}} \right|}^2}}  + {\sigma ^2}}}} \right)} }}{{\sum\nolimits_{k = 1}^K {P_k^{\left( l \right)}}  + {P_{c}}}}, \label{solve_eta}
	\end{flalign}
	where $P_k^{\left(l\right)}$ denotes the local point at the $l$-th iteration.
	
	Although the feasible set of problem (\ref{p3}) is convex since it is characterized by multiple linear constraints, the objective function remains non-concave due to the non-convex term ${\log _2}\left( {1 + \frac{{{P_k}{{\left| {{\bf{w}}_k^H{{\bf{h}}_k}} \right|}^2}}}{{\sum\nolimits_{i = k+1}^K {{P_i}{{\left| {{\bf{w}}_k^H{{\bf{h}}_i}} \right|}^2}}  + {\sigma ^2}}}} \right)$. To address this issue, we first recast it into a difference of convex (DC) form as
	\begin{flalign}
		&{{{\log }_2}\left( {{P_k}{{\left| {{\bf{w}}_k^H{{\bf{h}}_k}} \right|}^2} + \sum\nolimits_{i = k + 1}^K {{P_i}\left| {{\bf{w}}_k^H{{\bf{h}}_i}} \right|^2}  + {\sigma ^2}} \right)} \notag\\
		&~~~{ - {{\log }_2}\left( {\sum\nolimits_{i = k + 1}^K {{P_i}\left| {{\bf{w}}_k^H{{\bf{h}}_i}} \right|^2}  + {\sigma ^2}} \right)} .
	\end{flalign}
	Next, the concave function ${{\log }_2}\left( {\sum\nolimits_{i = k + 1}^K {{P_i}\left| {{\bf{w}}_k^H{{\bf{h}}_i}} \right|^2}  + {\sigma ^2}} \right)$ is replaced by its first-order Taylor upper bound, which is shown at eq. (\ref{A}).
	\begin{figure*}[!t]
		\begin{flalign}
			{\log _2}\left( {\sum\limits_{i = k + 1}^K {{P_i}{{\left| {{\bf{w}}_k^H{{\bf{h}}_i}} \right|}^2}}  + {\sigma ^2}} \right) \le {\log _2}\left( {\sum\limits_{i = k + 1}^K {P_i^{\left( l \right)}{{\left| {{\bf{w}}_k^H{{\bf{h}}_i}} \right|}^2}}  + {\sigma ^2}} \right) + \frac{{\sum\nolimits_{i = k + 1}^K {{{\left| {{\bf{w}}_k^H{{\bf{h}}_i}} \right|}^2}\left( {{P_i} - P_i^{\left( l \right)}} \right)} }}{{\left( {\sum\nolimits_{i = k + 1}^K {P_i^{\left( l \right)}{{\left| {{\bf{w}}_k^H{{\bf{h}}_i}} \right|}^2}}  + {\sigma ^2}} \right)\ln 2}} \buildrel \Delta \over = {\Lambda _k}, \label{A}
		\end{flalign}
		%\hrulefill
	\end{figure*}
	Accordingly, problem (\ref{p3}) can be recast as
	\begin{subequations}\label{p4}
		\begin{align}
			{\rm{  }}\mathop {{\rm{max}}}\limits_{\left\{P_k\right\}} &\sum\limits_{k = 1}^K {\left( {{{\log }_2}\left( { \sum\nolimits_{i = k}^K {{P_i}\left| {{\bf{w}}_k^H{{\bf{h}}_i}} \right|^2}  + {\sigma ^2}} \right)} - {\Lambda_k} \right)} \notag\\
			&~~~~~~~~- \eta  {\left(\sum\nolimits_{k = 1}^K {{P_k}}  + {P_{c}} \right) } \\
			&{\rm{s}}{\rm{.t}}{\rm{.}}~{\rm \left(\ref{p3-b}\right)}.
		\end{align}
	\end{subequations}
	
	As a result, the transformed problem (\ref{p4}) is convex, with a concave objective function and a convex feasible set. Accordingly, an FP-SCA algorithm is developed to obtain high-quality power allocation coefficients, and the overall procedure is summarized in {\bf Algorithm \ref{alg1}}.
	\begin{algorithm}[!t]
		\caption{FP-SCA Algorithm for Power Allocation}
		\label{alg1}
		\begin{algorithmic}[1]
			\STATE Initialize $\eta ^{\left(0\right)}$ and $\big\{P_k^{\left(0\right)}\big\}$, set iteration index $l = 0$.
			\REPEAT
			\STATE $l = l + 1$.
			\STATE Given $\eta ^{\left(l-1\right)}$, update $\big\{P_k^{\left(l\right)}\big\}$ by solving problem (\ref{p4}).
			\STATE Given $\big\{P_k^{\left(l\right)}\big\}$, update $\eta ^{\left(l\right)}$ by calculating eq. (\ref{solve_eta}).
			\UNTIL Convergence.
		\end{algorithmic}
	\end{algorithm}
	
	\subsection{Rotation Optimization}
	Given $\left\{{\bf w}_k\right\}$ and $\left\{{P_k}\right\}$, problem (\ref{p1}) is reformulated as
	\begin{subequations}\label{p5}
		\begin{align}
			{\rm{  }}\mathop {{\rm{max}}}\limits_{\left\{ {{\vec{\bf{f}}_n}} \right\}} ~&{\bar P}{\sum\limits_{k = 1}^K {{\log _2}\left( {1 + \frac{{{P_k}{{\left| {{\bf{w}}_k^H{{\bf{h}}_k}} \right|}^2}}}{{\sum\nolimits_{i = k+1}^K {{P_i}{{\left| {{\bf{w}}_k^H{{\bf{h}}_i}} \right|}^2}}  + {\sigma ^2}}}} \right)} } \\
			{\rm{s}}{\rm{.t}}{\rm{.}}~&R_k \ge {R_{\min }},\forall k,~\left( {\rm \ref{c_f}} \right),
		\end{align}
	\end{subequations}
	where ${\bar P} = \left({\sum\nolimits_{k = 1}^K {P_k + P_{c}}}\right)^{-1}$.
	To further deal with problem (\ref{p5}), we set ${h_{n,k}}\left( {{\vec{\bf{f}}_n}} \right) = {\bar{g}_{n,k}}{\left( {\vec{\bf{f}}_n^T{\vec{\bf{q}}_{n,k}}} \right)^p}$ and ${{\bf{h}}_k}\left( {{\bf{F}}} \right) = {\left[ {{h_{1,k}}\left( {{\vec{\bf{f}}_1}} \right),...,{h_{N,k}}\left( {{\vec{\bf{f}}_N}} \right)} \right]^T}$. Accordingly, problem (\ref{p5}) can be recast as
	\begin{subequations}\label{p6}
		\begin{align}
			{\rm{  }}\mathop {{\rm{max}}}\limits_{\left\{ {{\vec{\bf{f}}_n}} \right\}} ~&{\bar P}{\sum\limits_{k = 1}^K {{{{\log }_2}\left( {1 + \frac{{{P_k}{{\left| {{\bf{w}}_k^H{{\bf{h}}_k}\left( {\bf{F}} \right)} \right|}^2}}}{{\sum\nolimits_{i = k + 1}^K {{P_i}{{\left| {{\bf{w}}_k^H{{\bf{h}}_i}\left( {\bf{F}} \right)} \right|}^2}}  + {\sigma ^2}}}} \right)}} } \\
			{\rm{s}}{\rm{.t}}{\rm{.}}~&{P_k}{\left| {{\bf{w}}_k^H{{\bf{h}}_k\left(\bf F\right)}} \right|^2} \ge \left( {{2^{{R_{\min }}}} - 1} \right) \notag\\
			&\left( {\sum\nolimits_{i = k + 1}^K {{P_i}\left| {{\bf{w}}_k^H{{\bf{h}}_i\left(\bf F\right)}} \right|^2}  + {\sigma ^2}} \right),\forall k,~\left( {\rm \ref{c_f}} \right),
		\end{align}
	\end{subequations}
	
	By exploiting the DC transformation, the objective function of problem (\ref{p6}) can be equivalently rewritten as
	\begin{flalign}
		&{{{\log }_2}\left( {\sum\nolimits_{i = k}^K {{P_i}\left| {{\bf{w}}_k^H{{\bf{h}}_i\left(\bf F\right)}} \right|^2}  + {\sigma ^2}} \right)} \notag\\
		&~~~~~~{ - {{\log }_2}\left( {\sum\nolimits_{i = k + 1}^K {{P_i}\left| {{\bf{w}}_k^H{{\bf{h}}_i\left(\bf F\right)}} \right|^2}  + {\sigma ^2}} \right)} .
	\end{flalign}
	
	However, due to the presence of the convex term ${{{\left| {{\bf{w}}_k^H{{\bf{h}}_k}\left( {{\bf{F}}} \right)} \right|}^2}}$, the objective function and the QoS constraints in problem (\ref{p6}) are still non-convex. To address this issue, we approximate ${{{\left| {{\bf{w}}_k^H{{\bf{h}}_i}\left( {{\bf{F}}} \right)} \right|}^2}}\buildrel \Delta \over = {\Gamma _{k,i}}$ and ${\log _2}\left( {\sum\nolimits_{i = k + 1}^K {{P_i}{{\left| {{\bf{w}}_k^H{{\bf{h}}_i}\left( {{{\bf{F}}^{\left( x \right)}}} \right)} \right|}^2}}  + {\sigma ^2}} \right) \buildrel \Delta \over = {\Upsilon _k}$ by a first-order Taylor lower bound and a first-order Taylor upper bound, respectively, as shown in eqs. (\ref{F}) and (\ref{R}), where $\bar {\bf{h}} _{n,k}^{\left( x \right)} \buildrel \Delta \over = \frac{{\partial {h_{n,k}}\left( {\vec {\bf{f}} _n^{\left( x \right)}} \right)}}{{\partial \vec {\bf{f}} _n^{\left( x \right)}}} = {\bar{g} _{n,k}}p{\big( {{{\big( {\vec {\bf{f}} _n^{\left( x \right)}} \big)}^T}{{\vec {\bf{q}} }_{n,k}}} \big)^{p - 1}}{\vec {\bf{q}} _{n,k}}$.
	\begin{figure*}[!t]
		\begin{flalign}
			&{\Gamma _{k,i}} \buildrel \Delta \over = {\left| {{\bf{w}}_k^H{\bf{h}}_i^{\left( x \right)}\left( {{\bf{F}}} \right)} \right|^2} + \Re \left\{ {2{{\left( {{\bf{w}}_k^H{\bf{h}}_i^{\left( x \right)}\left( {{\bf{F}}} \right)} \right)}^*}\sum\nolimits_{n = 1}^N {{w_{n,k}^*}{{\left( {\bar {\bf{h}} _{n,i}^{\left( x \right)}} \right)}^T}\left( {{\vec{\bf{f}}_n} - \vec{\bf{f}}_n^{\left( x \right)}} \right)} } \right\},  \label{F} \\
			&{\Upsilon _k} \buildrel \Delta \over = {\log _2}\left( {\sum\limits_{i = k + 1}^K {{P_i}{{\left| {{\bf{w}}_k^H{{\bf{h}}_i}\left( {{{\bf{F}}^{\left( x \right)}}} \right)} \right|}^2}}  + {\sigma ^2}} \right) + \frac{{\sum\limits_{i = k + 1}^K {\Re \left\{ {2{P_i}{{\left( {{\bf{w}}_k^H{{\bf{h}}_i}\left( {{{\bf{F}}^{\left( x \right)}}} \right)} \right)}^*}\sum\nolimits_{n = 1}^N {w_{n,k}^*{{\left( {\bar {\bf{h}} _{n,i}^{\left( x \right)}} \right)}^T}\left( {{{\vec {\bf{f}} }_n} - \vec {\bf{f}} _n^{\left( x \right)}} \right)} } \right\}} }}{{\left( {\sum\nolimits_{i = k + 1}^K {{P_i}{{\left| {{\bf{w}}_k^H{{\bf{h}}_i}\left( {{{\bf{F}}^{\left( x \right)}}} \right)} \right|}^2}}  + {\sigma ^2}} \right)\ln 2}}. \label{R}
		\end{flalign}
		\hrulefill
	\end{figure*}
	In addition, the non-convex constraint (\ref{c_f}) is transformed into a convex form by applying a relaxation and trigonometric reformulation as
	\begin{flalign}
		\big\| {{\vec{\bf{f}}_n}} \big\| \le 1,\cos \left( {{\theta _{\max }}} \right) \le \vec{\bf{f}}_n^T{\bf{e}} \le 1,\forall n. \label{newf}
	\end{flalign}
	
	To avoid the nested exponentiation of optimization variables and facilitate convex optimization, we introduce an auxiliary variable $\xi _{n,k}$ to replace ${\big( {\vec{\bf{f}}_n^T{\vec{\bf{q}}_{n,k}}} \big)^{2p}}$. By the Cauchy-Schwarz inequality, an affine upper bound is given by ${\left| {{\bf{w}}_k^H{{\bf{h}}_i}\left( {{\bf{F}}} \right)} \right|^2} \le \big\|{\rm{diag}}\big(\bar{\bf g}_i^*\big){\bf w}_k\big\|^2\sum\nolimits_{n=1}^N\xi_{n,i} \buildrel \Delta \over = {\chi _{k,i} }$. After the above transformations, problem (\ref{p6}) is recast as
	\begin{subequations}\label{p7}
		\begin{align}
			&{\rm{  }}\mathop {{\rm{max}}}\limits_{\left\{ {{\vec{\bf{f}}_n}} \right\},\left\{{\bm \xi}_k\right\}} ~{\bar P}{\sum\limits_{k = 1}^K \left({{{\log }_2}\left( {\sum\nolimits_{i = k}^K {{P_i}{\Gamma_{k,i}}}  + {\sigma ^2}} \right) - {\Upsilon _k}}\right) } \\
			&{\rm{s}}{\rm{.t}}{\rm{.}}~{P_k}{\Gamma_{k,k}} \ge \left( {{2^{{R_{\min }}}} - 1} \right) \left( {\sum\nolimits_{i = k + 1}^K {{P_i}{\chi_{k,i}}}  + {\sigma ^2}} \right),\forall k,\notag\\
			&~~~~~{\xi _{n,k}} \ge {\left( {\vec {\bf{f}} _n^T{{\vec {\bf{q}} }_{n,k}}} \right)^{2p}}, \forall n,k, ~(\rm \ref{newf}).
		\end{align}
	\end{subequations}
	
	As a result, the objective function of problem (\ref{p7}) becomes concave, and the feasible set is convex. Therefore, this problem has been converted into a convex problem, which can be solved directly by using CVX \cite{grant2014cvx}. Next, the obtained $\vec{\bf f}_n $ is normalized to recover a unit-norm pointing vector, i.e., $\vec{\bf f}_n = \vec{\bf f}_n \big\|\vec{\bf f}_n \big\|^{-1}$.
	
	Finally, the three subproblems are alternately solved until the convergence criterion is satisfied. The complete steps are presented in {\bf Algorithm \ref{alg2}}. The proposed algorithm has an approximate computational complexity of ${\cal O}\big( {{I}\big( K + LK^{3.5} + K^{0.5}(3+K)^3 N^{3.5} \big)} \big)$, where $L$ and $I$ are the numbers of iterations required by {\bf Algorithms \ref{alg1}} and {\bf \ref{alg2}}, respectively. The empirical convergence behavior is verified in Section IV.
	\begin{algorithm}[!t]
		\caption{BCD-Based Algorithm for Problem (\ref{p1})}
		\label{alg2}
		\begin{algorithmic}[1]
			\STATE Initialize $\big\{ {{{\bf{w}}_k^{\left(0\right)}}} \big\}$, $\big\{ P_k^{\left(0\right)} \big\}$, $\big\{ {{\vec{\bf{f}}_n^{\left(0\right)}}} \big\}$, set $x = 0$.
			\REPEAT
			\STATE $x = x + 1$.
			\STATE Given $\big\{ P_k^{\left(x-1\right)} \big\}$ and $\big\{ {{\vec{\bf{f}}_n^{\left(x-1\right)}}} \big\}$, update $\big\{ {{{\bf{w}}_k^{\left(x\right)}}} \big\}$ by (\ref{solvew}).
			\STATE Given $\big\{ {{{\bf{w}}_k^{\left(x\right)}}} \big\}$ and $\big\{ {{\vec{\bf{f}}_n^{\left(x-1\right)}}} \big\}$, update $\big\{ P_k^{\left(x\right)} \big\}$ by {\bf Algorithm \ref{alg1}}.
			\STATE Given $\big\{ {{{\bf{w}}_k^{\left(x\right)}}} \big\}$ and $\big\{ P_k^{\left(x\right)} \big\}$, update $\big\{ {{\vec{\bf{f}}_n^{\left(x\right)}}} \big\}$ by solving problem (\ref{p7}).
			\UNTIL Convergence.
		\end{algorithmic}
	\end{algorithm}

	\section{Numerical Results}	
	Following \cite{11427014,11098736,11347570}, the simulation parameters are set as follows: $K_G = K_A = 2$, $N_x = N_y = 3$, $P_{\max} = 1~{\rm mW}$, $\lambda = 0.125$, $\sigma^2 = -80~{\rm dBm}$, $R_{\min} = 0.2~{\rm bps/Hz}$, $p = 4$, $\theta_{\max} = 60^{\circ}$, and $H_{\rm BS} = 5~{\rm m}$. All users are uniformly distributed within a $90^\circ$ sector centered at the BS, where the radial distance from the BS ranges from $50~{\rm m}$ to $60~{\rm m}$. The aerial users are assumed to fly at an altitude of $20~{\rm m}$. To benchmark the proposed RA-NOMA scheme, we consider two multiple-access counterparts employing SDMA and TDMA protocols, termed as RA-SDMA and RA-TDMA, respectively. In addition, the IA and the FA schemes are included, denoted by IA-NOMA and FA-NOMA, respectively.
	
\begin{figure}[!t]
		\centering
		\begin{subfigure}{0.49\linewidth}
			\centering
			\includegraphics[width=\linewidth]{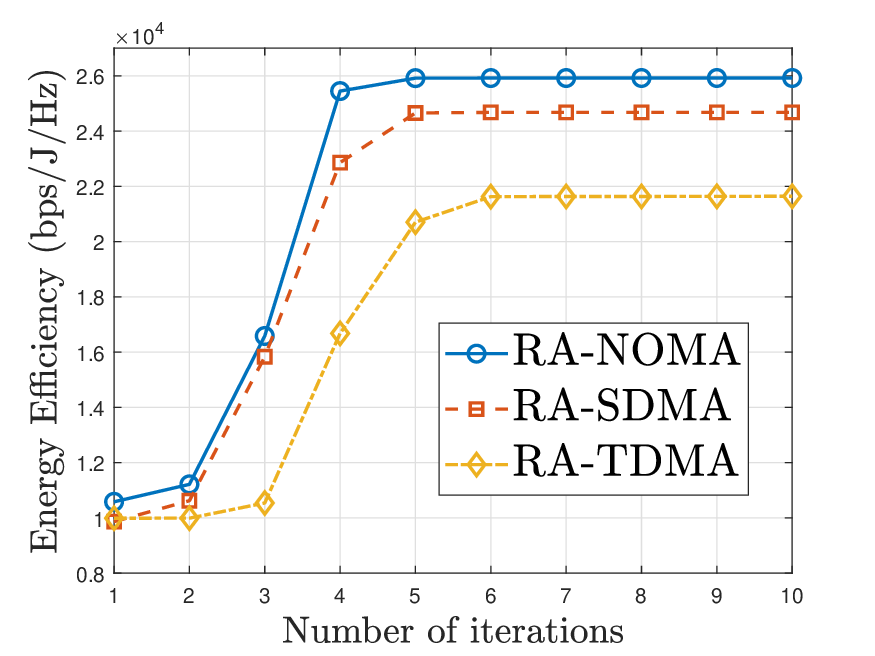}
			\subcaption{\centering Algorithm 2.}
			\label{outer}
		\end{subfigure}
		\hfill
		\begin{subfigure}{0.49\linewidth}
			\centering
			\includegraphics[width=\linewidth]{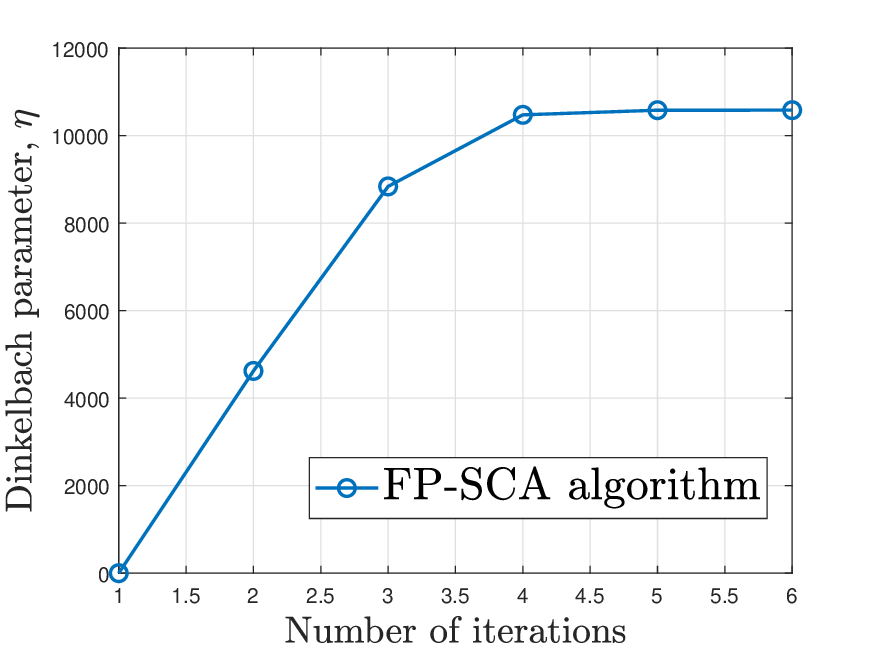}
			\subcaption{\centering Algorithm 1.}
			\label{inner}
		\end{subfigure}
		\caption{Convergence behavior of the proposed algorithms.}
		\label{iter}
	\end{figure}

	Figs. \ref{iter}(\subref{outer}) and (\subref{inner}) verify the favorable convergence behavior of the proposed {\bf Algorithms \ref{alg1}} and {\bf \ref{alg2}}, respectively. As shown in Fig. \ref{iter}(\subref{outer}), the RA-NOMA, RA-SDMA, and RA-TDMA algorithms all converge within a few iterations. RA-TDMA converges slightly slower due to the additional time allocation optimization. Overall, the results verify the reliable convergence of the proposed algorithms.
	
\begin{figure}[!t]
		\centering
		\includegraphics[width=2.4in]{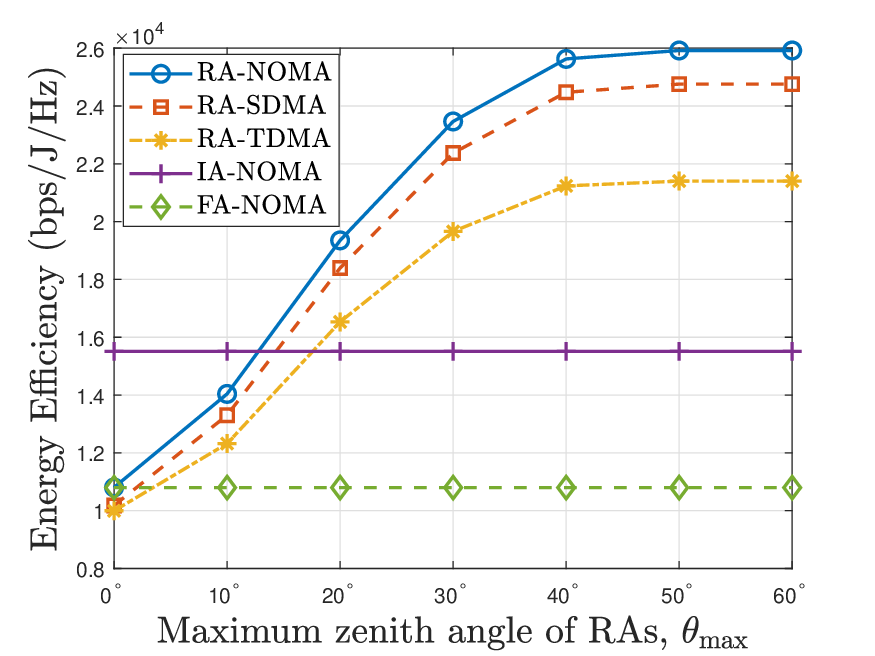}
		\caption{Maximum zenith angle versus energy efficiency.}
		\label{theta}
	\end{figure}

	Fig. \ref{theta} illustrates the impact of the maximum zenith angle $\theta_{\max}$ on the system EE. It can be observed that the EE increases with $\theta_{\max}$, since a larger maximum zenith angle provides the RA system with higher spatial flexibility for adapting its pointing directions. When $\theta_{\max}$ exceeds $50^\circ$, the EE tends to saturate, indicating that such an angular range is already sufficient to cover all users in the considered scenario. Moreover, the RA-NOMA scheme achieves a significantly higher EE than the RA-SDMA and RA-TDMA schemes, which can be attributed to the high spectral utilization of NOMA and its interference cancellation capability enabled by SIC. In addition, the IA-NOMA scheme outperforms the FA-NOMA scheme owing to the omnidirectional signal reception capability of isotropic antennas. 
	
\begin{figure}[!t]
		\centering
		\includegraphics[width=2.4in]{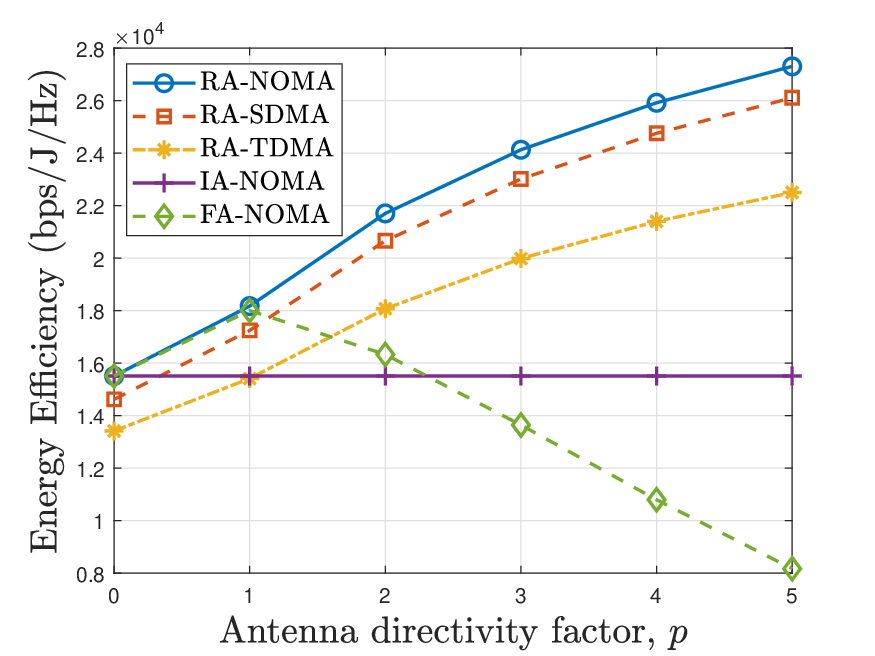}
		\caption{Antenna directivity factor versus energy efficiency.}
		\label{p}
	\end{figure}

	Fig. \ref{p} presents the system EE versus the antenna directivity factor $p$. Here, the EE of the RA-based schemes steadily improves as $p$ increases. Because a larger $p$ leads to a more directive radiation pattern, enabling the RA system to achieve more accurate beam pointing and stronger directional gains. In contrast, the EE of the FA-NOMA scheme first increases and then decreases with $p$. Although increasing $p$ enhances the beam directivity, the fixed antenna orientations cannot adapt to user locations. As a result, highly directive beams are more likely to deviate from the user directions, thereby reducing the effective directional gain.
	
\begin{figure}[!t]
		\centering
		\includegraphics[width=2.4in]{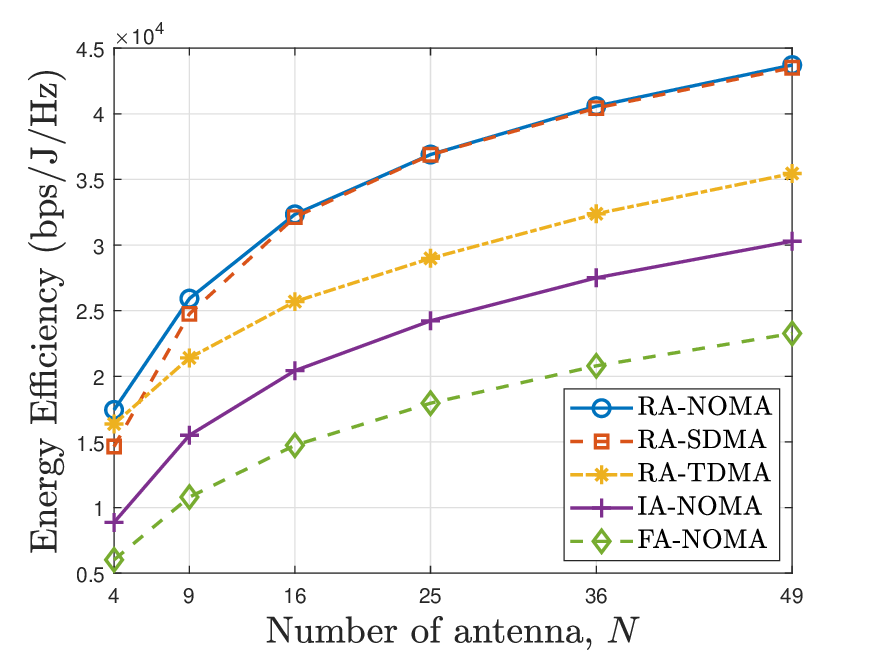}
		\caption{Number of antennas versus energy efficiency.}
		\label{N}
	\end{figure}

	Fig. \ref{N} presents the system EE under different numbers of antennas $N$. In fig. \ref{N}, increasing $N$ leads to a continuous improvement in EE, which can be attributed to the higher received signal strength and stronger receive beamforming capability brought by a larger antenna array. Moreover, the EE difference between the RA-NOMA and RA-SDMA schemes decreases as $N$ grows. This phenomenon occurs because additional antennas offer more spatial DoFs, allowing SDMA to suppress inter-user interference more effectively via receive beamforming. By comparison, the advantage of RA-NOMA over RA-TDMA becomes more evident with a larger $N$, since NOMA enables multiple users to transmit over the same time-frequency resources, while TDMA inevitably incurs performance loss due to time-slot division.

	\section{Conclusion}
	This paper investigated a RA-assisted uplink NOMA system for energy-efficient information reception from ground and aerial users. By jointly optimizing the BS receive beamforming, user transmit power allocation, and RA rotation angles, an EE maximization framework was developed. The non-convex problem was addressed by combining MMSE-based receive beamforming, FP-SCA-based power allocation, and convexified RA rotation optimization. Simulation results confirmed the advantages of the proposed RA-NOMA scheme over RA-SDMA, RA-TDMA, IA-NOMA, and FA-NOMA benchmarks, demonstrating the potential of RA for improving energy-efficient NOMA transmission.

	\bibliography{Reference-NOMA}
	\bibliographystyle{IEEEtran}
	
\end{document}